\documentclass[conference]{IEEEtran}
\usepackage{slashbox}
\usepackage{amsmath}    
\usepackage{graphicx}   
\usepackage{graphics}   
\usepackage{verbatim}   
\usepackage{color}      
\usepackage{subfigure}  
\begin{document}
%
\title{Channel Identification and its Impact on Quantum LDPC Code Performance}


\author{\IEEEauthorblockN{Yixuan Xie, Jun Li, Robert Malaney and Jinhong Yuan}\\
\IEEEauthorblockA{School of Electrical Engineering and Telecommunications\\
The University of New South Wales, Sydney, Australia\\
Email: Yixuan.Xie@student.unsw.edu.au, Jun.Li@unsw.edu.au,
R.Malaney@unsw.edu.au, J.Yuan@unsw.edu.au}} \maketitle
\maketitle

\begin{abstract}
In this work we probe the impact of channel estimation on the performance of quantum LDPC codes.  Our channel estimation is based on an optimal estimate of the relevant decoherence parameter via  its quantum Fisher information. Using state-of-the art quantum LDPC codes designed for the  quantum depolarization channel, and utilizing  various quantum probes with different entanglement properties, we show how the performance of such codes can deteriorate by an order of magnitude when optimal channel identification is fed into a belief propagation decoding algorithm. Our work highlights the importance in quantum communications of a viable channel identification campaign prior to decoding, and highlights the trade-off between entanglement  consumption and quantum LDPC code performance.
\end{abstract}


%
\IEEEpeerreviewmaketitle

\section{Introduction}

A quantum error correction code (QECC) is intended to protect  fragile quantum states from unwanted evolution and allow for robust implementations of quantum processing devices. Compared to classical communication systems, error correction in quantum communication channels is challenging, since a quantum bit, namely a \textit{qubit}, has continuous states rather than two discrete states. Implementation of practical encoders and decoders in quantum codes is further complicated by the fact that a quantum state is usually affected by measurement, and that an unknown quantum  state cannot be duplicated.

For classical error correction, it is well known that practical decodable codes exist. When an optimal decoder is applied, classical codes can achieve information rates close to the Shannon limit. Low-density parity-check (LDPC) codes \cite{Gallager:19633} \cite{MN1996} are an example of such codes. The sparseness of the parity-check matrices makes the codes easy to encode and decode, even when communicating very close to the Shannon limit. It is established that the sum-product message passing algorithm is an optimal decoding algorithm for LDPC codes provided that the factor graph of LDPC codes is a tree structure, \emph{i.e.} no cycles exist.

Following the discovery of CSS (Calderbank, Shor and Steane) codes \cite{Calderbank1996} \cite{Steane} and stabilizer codes \cite{Gottesman1996}, it has been known how quantum error-correction codes can be developed in a similar manner to classical codes. Quantum LDPC codes based on finite geometry were first proposed in \cite{Postol}. However, a key  constraint on the matrix representing the stabilizers (arising from commutativity requirements) makes the design of quantum LDPC codes difficult. In \cite{MacKay2004}, Mackay \textit{et al.} proposed the \emph{bicycle} codes and explored the conjecture that the best quantum error-correcting codes will be closely related to the best classical codes, and in \cite{poulinturbo} Poulin \textit{et al.} proposed serial turbo codes for quantum error correction.  A more detailed history on the development of QECC can be found elsewhere \emph{e.g.} \cite{Nielsen:2000}, \cite{Knill1997}. More recently, many works attempting to improve quantum LDPC code performance have been published, and in this regard  the recent codes  of \cite{Tan:2010}, \cite{Hagiwara:2011}, and \cite{kenta2011}  based on quasi-cyclic structure  can be considered as representative of state-of-the-art quantum LDPC codes.



Hitherto, in investigations of the performance of  quantum LDPC codes it has been assumed that perfect knowledge of the quantum channel exists. Of course in practice this is not the case. In this work we probe the impact of imperfect channel knowledge on the performance of quantum LDPC codes. We will utilize optimal estimates of the channel derived from quantum Fisher information about the channel parameters. More specifically, we will investigate the performance of quantum codes over the depolarization channel when optimal estimates of the depolarizing parameter are available.

In section II we first briefly review quantum communications and the \textit{stabilizer} formalism for describing QECCs, and discuss their relationship to classical codes. In section III we review quantum channels and quantum channel identification. In section IV we present our simulation results using existing quantum stabilizer codes from \cite{Tan:2010} over the commonly used quantum \textit{Depolarizing Channel}, showing how the imperfect channel identification impacts the codes performance. Lastly, we draw some conclusions and discuss future works.

\section{Quantum Codes}
The analog of classical `bit' is a `qubit', which can be represented as a quantum state \textit{$\left|\psi\right\rangle$} in a two-dimensional complex vector space. This can be written as a \textit{superposition}
\begin{equation}
\label{eq1}
\left|\psi\right\rangle = \alpha_{0}\left|0\right\rangle + \alpha_{1}\left|1\right\rangle
\end{equation}
where $\alpha_{0}$ and $\alpha_{1}$ are complex numbers satisfying $|\alpha_{0}|^{2} + |\alpha_{1}|^{2} = 1$. The quantum state of \textit{N} qubits has the form $\sum{\alpha_{s}\left|s\right\rangle}$, where $s$ runs over all binary strings of length \textit{N}. The $2^{\textit{N}}$ independent complex coefficients $\alpha_{s}$ then satisfy the normalization constraint $\left|\alpha_{0}\right|^{2} + \left|\alpha_{1}\right|^{2} + \ldots + \left|\alpha_{2^{N}-1}\right|^{2} = 1$. Each of the $2^{N}$ states $\left|000\ldots0\right\rangle$,$\left|000\ldots1\right\rangle$ \ldots $\left|111\ldots1\right\rangle$ is shorthand for the \textit{N-fold} tensor product $\left|0\right\rangle\otimes\left|0\right\rangle\ldots\otimes\left|0\right\rangle$, $\left|0\right\rangle\otimes\left|0\right\rangle\ldots\otimes\left|1\right\rangle$, \ldots $\left|1\right\rangle\otimes\left|1\right\rangle\ldots\otimes\left|1\right\rangle$. Suppose a quantum state $\left|\psi\right\rangle$ of size $N$ is sent through a quantum noisy channel. The outcome of the transmission can be written as $E\left|\psi\right\rangle$, where the error operator $E$ takes the form of $E = e_{1}\otimes e_{2}\otimes \ldots \otimes e_{N}$, which can be considered as an \textit{N-fold} tensor product of errors operators $e_j$, where $j = 1 \ldots N$.

A typical quantum channel is the \emph{Pauli Channel} in which  the error operators can be modeled by the three different Pauli operators
\begin{equation*}
\label{PauliOperators}
X = \left( {\begin{array}{*{20}{c}}
0&1\\
1&0
\end{array}} \right),Z = \left( {\begin{array}{*{20}{c}}
1&0\\
0&{ - 1}
\end{array}} \right),Y = \left( {\begin{array}{*{20}{c}}
0&{ - i}\\
i&0
\end{array}} \right),
\end{equation*}
and the $2 \times 2$ identity matrix $I$. These matrices form a Pauli group \texttt{$P_{N}$} that act on $N$ qubits. The elements of \texttt{$P_{N}$} either commute or anti-commute, and for all error operators $E$, $F$ $\in$ \texttt{$P_{N}$}, the commutativity between them is defined as
\begin{equation*}
E \circ F = \left\{ {\begin{array}{*{20}{l}}
1&{if}&{EF = FE}\\
{ - 1}&{if}&{EF =  - FE}
\end{array}} \right..
\label{commutativity}
\end{equation*}
In what follows we will also refer to the Pauli matrices $I$, $X$, $Y$, and $Z$, as  $\sigma_0$, $\sigma_1$, $\sigma_2$ and $\sigma_3$, respectively.
%

\subsection{Preliminary on Quantum Stabilizer Codes}
\label{DefStabiler}
A stabilizer generator $S$ that encodes $K$ qubits in $N$ qubits consists of a set of Pauli operators on the $N$ qubits closed under multiplication, with the property that any two operators in the set \textit{commute}, so that every stabilizer can be measured simultaneously. An example of a stabilizer generator $S$ is shown below for $K=1, N=5$ representing a rate $\frac{1}{5}$ quantum stabilizer code,
\begin{equation}
\label{stabilizer}
S = \left(\begin{array}{ccccc}
Z & Z & X & I & X \\
X & Z & Z & X & I \\
I & X & Z & Z & X \\
X & I & X & Z & Z
\end{array}\right).
\end{equation}

Consider now a set of error operators $\left\{E_{\alpha}\right\}$ taking a state $\left|\psi\right\rangle$ to the corrupted state $E_{\alpha}\left|\psi\right\rangle$. A given error operator either commutes or anti-commutes with each stabilizer $S_{i}$ (row of the generator $S$) where $i=1 \ldots N-K$. If the error operator commutes with $S_{i}$ then
\begin{equation}
\label{errorcommu}
S_{i}E_{\alpha}\left|\psi\right\rangle = E_{\alpha}S_{i}\left|\psi\right\rangle = E_{\alpha}\left|\psi\right\rangle
\end{equation}
 and therefore $E_{\alpha}\left|\psi\right\rangle$ is a $+1$ eigenstate of $S_{i}$. Similarly, if it anti-commutes with $S_{i}$, the eigenstate is $-1.$ The measurement outcome of $E_{\alpha}\left|\psi\right\rangle$ is known as the \textit{syndrome}.

\subsection{Conversion between Quantum and Classical Codes}

To connect quantum stabilizer codes with classical LDPC codes it is useful to describe any given Pauli operator on $N$ qubits as a product of an $X$-containing operator, a $Z$-containing operator and a phase factor $\left(+1,-1,i,-i\right)$. For example, the first row of matrix (\ref{stabilizer}) can be expressed as

\begin{align}
ZZXIX = (IIXIX) \times (ZZIII).
\end{align}
Thus, we can directly express the $X$-containing operator and $Z$-containing operator as separate binary strings of length $N$. In the  $X$-containing operator a $1$ represents the  $X$ operator (likewise for the $Z$ operator), and $0$ for $I$. The resulting binary formalism of the stabilizer is a matrix $ A = \left(A_{1} | A_{2}\right)$ of $2N$ columns and $M = N-K$ rows,  where $A_{1}$ and $A_{2}$ represent $X$-containing and $Z$-containing operators, respectively
\newline
\emph{Example 1:} For example, the set of stabilizers in (\ref{stabilizer}) appears as the binary matrix $A$

\begin{equation}
\centering
\label{binaryStabilizer}
A = \left(A_{1} | A_{2}\right) = \left(\begin{array}{c|c}
\begin{array}{ccccc}
0 & 0 &  1 & 0 & 1 \\
1 & 0 &  0 & 1 & 0 \\
0 & 1 &  0 & 0 & 1 \\
1 & 0 &  1 & 0 & 0 \\
\end{array}
&
\begin{array}{ccccc}
1 &  1 & 0 & 0 & 0 \\
0 &  1 & 1 & 0 & 0\\
0 &  0 & 1 & 1 & 0\\
0 &  0 & 0 & 1 & 1\\
\end{array}
\end{array}
\right).
\end{equation}

%

Due to the requirement that stabilizers must commute, a constraint  on a general matrix $A$ can be written as  \emph{e.g.} \cite{MacKay2004}.

\begin{equation}
\centering
\label{binaryConstraint}
A_{1}A_{2}^{T}+A_{2}A_{1}^{T} = 0.
\end{equation}
 Note that the quantum syndrome can be conceptually considered as an equivalent to the classical syndrome $Ae$, where $A$ is a binary parity-check matrix and $e$ is a binary error vector.

To summarize, the property of stabilizer codes can be directly inferred from classical codes. Any binary parity-check matrix of size $M\times 2N$ that satisfies the constraint in (\ref{binaryConstraint}) defines a quantum stabilizer code with rate $R = \frac{K}{N}$ that encodes $K$ qubits into $N$ qubits.

\subsection{CSS Codes}
As mentioned earlier, an important class of codes are the \textit{CSS Codes}  \cite{Calderbank1996}\cite{Steane}. These have the form

\begin{equation}
\label{CSS}
A = \left(\begin{array}{c|c}
\begin{array}{c}
H \\
0 \\
\end{array}
&
\begin{array}{c}
0 \\
G \\
\end{array}
\end{array}
\right)
\end{equation}
where $H$ and $G$ are $M_{H}\times N$ and $M_{G}\times N$ matrices, respectively,  ($M_{H}$ does not necessary equal to $M_{G}$). Requiring $HG^{T} = 0$ ensures that constraint (\ref{binaryConstraint}) is satisfied.
 If $G = H$, the resulting CSS code structure is called a \textit{dual-containing code}. Most classical (good) LDPC codes do not satisfy the constraint (\ref{binaryConstraint}).

\section{Quantum Channel Models and Estimation}

\subsection{Quantum Channel Models}

  Given some initial system state $\left| {{\Psi _s}} \right\rangle $, a decoherence model can be built by studying the time evolution of the system state's interaction with some external environment with initial state  $\left| {{\Psi _e}} \right\rangle $. Without loss of generality we can assume $\left| {{\Psi _s}} \right\rangle $ and $\left| {{\Psi _e}} \right\rangle $ are initially not entangled with each other.

 In terms of the  density operators ${\rho _s} = \left| {{\Psi _s}} \right\rangle \left\langle {{\Psi _s}} \right|$ and ${\rho _e} = \left| {{\Psi _e}} \right\rangle \left\langle {{\Psi _e}} \right|$, the initial state of the combined total system can be written
 as ${\rho _s} \otimes {\rho _e}$. The closed evolution of ${\rho _s} \otimes {\rho _e}$ can be described by a unitary $U$ via ${U}({\rho _s} \otimes {\rho _e})U^\dag $. To obtain the output system state, $\rho _s^{out}$, after some closed evolution $U$, we use $\rho _s^{out} \equiv \varepsilon \left( {{\rho _s}} \right) = {\rm{T}}{{\rm{r}}_e}\left[ {{U}({\rho _s} \otimes {\rho _e})U^\dag } \right]$
where Tr$_e$ is the partial trace over the environment's qubits. The channel $\rho _s^{out} \equiv \varepsilon \left( {{\rho _s}} \right)$ is a completely positive, trace preserving, map which provides the required evolution of ${\rho _s}$. It is possible to describe such maps directly using  an operator-sum representation,
\begin{equation}
 \varepsilon \left( {{\rho _s}} \right) = \sum\limits_{a = 1}^{N_o} {{K_a}} {\rho _s}K_a^\dag ,{\rm{ \ where \ }}\sum\limits_{a = 1}^{N_o} {K_a^\dag {K_a} = I}  , \label{eqKraus}
\end{equation}
and where ${K_{a = 1...N_o}}$ represent the so-called Kraus operators, with $N_o$ being the number of Kraus operators \cite{Kraus1983}.

 There are of course decoherence channels modeled on specific qubit-environment interactions  (e.g. see \cite{Nielsen:2000}). In this work we will consider only the depolarization channel.
 Let us introduce the  depolarization parameter, $f$,  of a qubit where $0 \le f \le 1$, with $f=1$ meaning complete depolarization and  $f=0$ meaning no depolarization.
 In terms of the Pauli matrices ${\sigma _i}$ (here $i=0,1,2,3)$,
the depolarization channel for a single qubit can be defined as $\varepsilon \left( {{\rho _s}} \right) = (1 - f){\rho _s} + f\frac{{{\sigma _o}}}{2}$. Using the relation ${\sigma _o} = \frac{1}{2}\left( {{\rho _s} + \sum\limits_{j = 1}^3 {{\sigma_j}{\rho _s}{\sigma_j}} } \right) $, we see that the Kraus operators for the depolarization channel can be written
${K_1} = \sqrt {1 - \frac{{3f}}{4}} {\sigma _o}$, ${K_2} = \sqrt {\frac{f}{4}} {\sigma _x}$, ${K_3} = \sqrt {\frac{f}{4}} {\sigma _y}$, and ${K_4} = \sqrt {\frac{f}{4}} {\sigma _z}$.  Note that it is also possible to parameterize the depolarization channel as $\varepsilon ({\rho _s}) = (1 - f'){\rho _s} + \frac{{f'}}{3}\sum\limits_{j = 1}^3 {{\sigma _j}} {\rho _s}{\sigma _j}$, where $f' = \frac{3}{4}f$. This latter form is more convenient for decoding purposes, and below we term $f'$ as the \emph{flip probability}.

\subsection{Quantum Channel Estimation}
\label{QCE}
The issue of quantum channel identification  (quantum process tomography) is of fundamental importance  for a range of practical quantum information processing problems (\emph{e.g.}  \cite{Nielsen:2000}). In the context of LDPC quantum error correction codes, it is normally assumed that the quantum channel is known perfectly in order for the code design to proceed. In reality of course, perfect knowledge of the quantum channel is not available - only some estimate of the channel is available. The key focus of this work is an investigation of this issue. To make progress we will assume a depolarization channel with some parameter $f$. However, we assume the true value of $f$ is unknown \emph{a priori}, and must first be measured via some channel identification procedure. This estimate of $f$ will then be used in the decoder in order to measure its performance relative to a decoder in which the true $f$ is utilized.

In general, quantum channel identification proceeds by inputting a known  quantum state $\sigma$ (the probe) into a quantum channel $\Gamma_p$ that is dependent on some parameter  $p$ (in our case $p=f$). By taking some quantum measurements on the output quantum state  $\Gamma_p(\sigma)$ which leads to some result $R$, we then hope to estimate $p$ . The input quantum state may be unentangled, entangled with an ancilla qubit (or qudit), or entangled with another probe. Multiple probes could be used, or the same probe can be recycled (\emph{i.e}. sent through the channel again). As can be imagined many  experimental schemes could be developed along these lines, and the performance of each scheme (\emph{i.e.} how well it estimates the true value of the parameter $p$) could be analyzed. However, in this study we will take a different tact. Here we will simply assume an experimental set-up is realized that obtains the information-theoretical \emph{optimal} performance.

Optimal channel identification via the use of the quantum Fisher information has been well studied in recent years, particularly in regard to the determination of the parameter $f$ of the depolarizing channel  (\emph{e.g.} \cite{Fujiwara2001}, \cite{Sasaki2002}, \cite{Fujiwara2003}, \cite{Frey2010}, \cite{Frey2011}). Defining ${\rho _f} = {\Gamma _f}(\sigma )$, the
 quantum Fisher information about $f$ can be written as
 \begin{align*}
 J(f) = J\left( {{\rho _f}} \right) = {\rm{tr}}\left[ {{\rho _f}} \right]L_f^2,
 \end{align*}
where ${L_f}$ is the symmetric logarithmic derivative defined implicitly by
\begin{align*}
 2{\partial _f}{\rho _f} = {L_f}{\rho _f} + {\rho _f}{L_f},
\end{align*}
and where  ${\partial _f}$ signifies partial differential w.r.t. $f$. With the quantum Fisher information in hand, the quantum Cramer-Rao bound can then be written as
 \[{\rm{mse}}\left[ {\hat f} \right] \ge {\left( {N_mJ(f)} \right)^{ - 1}}\]
where ${\rm{mse}}\left[ {\hat f} \right]$
is the mean square error of the unbiased estimator ${\hat f}$, and $N_m$ is the number of independent quantum measurements. In the simulations pursued here we will assume the channel is constant over the block length of the codeword, and unless otherwise stated we assume $N_m=1$. Further, we will assume two different cases for the quantum probe. In case $A$ we will assume the qubit probe is in a pure unentangled state, and as such \cite{Fujiwara2001} the quantum Fisher information about $f$ relevant to each codeword can be shown to be  $J\left( f \right) = {\left[ {f\left( {2 - f} \right)} \right]^{ - 1}}$. In case $B$ we adopt the scenario where one pair of maximally entangled qubit pairs is consumed per transmission of each codeword (one of the qubits traverses the channel). In this latter case the quantum Fisher information about $f$ relevant to each codeword can be shown to be
$J\left( f \right) = {\left[ {f\left( {\frac{4}{3} - f} \right)} \right]^{ - 1}}$ \cite{Fujiwara2001}. Similar expressions for qudit probes are available \cite{Frey2011}.

\section{Simulations}
%

From the discussion in Section II, a stabilizer generator can  be described in the binary form, $A = \left(A_{1}|A_{2}\right)$. It is also worthwhile to note that it can also be described in a quaternary form, where $A_{1}$ and $A_{2}$ are packed into a single matrix with elements $I$, $X$, $Y$, $Z$. Since a close link between $\{I,X,Y,Z\}$ and $\{0,1,\omega,\omega^{2}\}$ exists, where $\omega$ is the primary element in GF($4$), a quantum stabilizer code can be thought of as an analog to a GF($4$) classic code.
Thus, a Belief Propagation (BP) decoding algorithm in GF(4) \cite{Poulin2008} can be applied to quantum stabilizer codes.
For lower computational complexity, we applied BP-decoding in GF(2). However, the decoder was modified from the pure classical decoder  in order to break the  degeneracy problem of stabilizer codes.  To break the degeneracy, a heuristic method presented in \cite{Poulin2008} was adopted.


\subsection{Iterative BP-Decoding Algorithm}
\label{SPdecoder}
Given our previous discussions, the decoding algorithm applied in our simulations can be viewed as a variation of the standard BP message-passing decoding algorithm taking place in the binary field, with the decoder treating the depolarizing channel as two independent binary symmetric channels. The received values $r_{i}$ (here $i = 1,\ldots n$, where $n=2N$ is the classical block length) are either $0$ or $1$. These can be mapped to measurement outcomes $s \in \left\{1, -1\right\}^{M}$ (syndrome) of the received qubit sequence, and this syndrome is then used in error estimation and recovery. Assuming an initial quantum state representing a codeword, the initial probabilities $p_i$ for the $ith$ qubit  of the state undergoing an $X$, $Y$ or $Z$ error  are
\begin{equation}
\label{LLRBSC}
p_{i} = \left\{ {\begin{array}{*{20}{c}}
{{f^{'}}}&{for}&{X, \  Y, \ or  \ Z}\\
{1 - {f^{'}}}&{for}&{I}
\end{array}} \right.,
\end{equation}
where $f'$ is the flip probability known at the decoder.

The standard BP algorithm operates by sending messages along the edges of the Tanner graph. Let $u_{b_{i} \to c_{j}}$ and $u_{c_{j} \to b_{i}}$ denote the messages sent from bit node \textit{i} to check node \textit{j} and messages sent from check node \textit{j} to bit node \textit{i}, respectively. Also denote $\textit{N}(b_{i})$ as the number of neighbors of bit node $i$, and define $\textit{N}(c_{j})$ as the number of neighbors of check node $j$.

To initialize our algorithm, each qubit node sends out a message to all its neighbors equal to its initial probability value $p$ obtained according to equation (\ref{LLRBSC}). Upon reception of these messages, each check node sends out a message to its neighboring qubit node given by
\begin{equation}
\label{checktobit}
u_{{c_j} \to {b_i}} = \sum\limits_{{t_{1 \ldots n}} \in \{ t |t \  \circ \ {c_j}^T = {s_j})\} } \ \ {\prod\limits_{b_i' \in N({c_j})\backslash {b_i}} {u_{{b_{i'}} \to {c_j}}}}
\end{equation}
 where $N\left(c_{j}\right)\backslash b_{i}$ denotes all neighbors of check node $j$ except qubit node $i$, and the summation is over all possible error sequences $t_{ 1\ldots N}$. Each bit node then sends out a message to its neighboring checks given by


\begin{equation}
\label{bittocheck}
u_{{b_i} \to {c_j}} = p_i\prod\limits_{{c_{j'}} \in N({b_i})\backslash {c_j}} {u_{{c_{j'}} \to {b_i}}}
\end{equation}
where $N\left(b_{i}\right)\backslash c_{j}$ denotes all neighbors of qubit node $i$ except check node $j$. Equations (\ref{checktobit}) and (\ref{bittocheck}) operate iteratively until the message is correctly decoded or the maximum pre-determined iteration number is reached.

%



\begin{figure}[htp]
\centering
\includegraphics[width=3.2in]{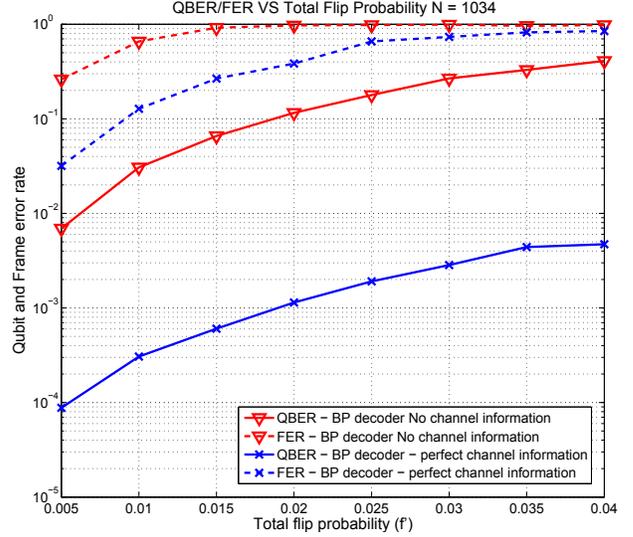}
\vspace{-0.3cm}
\caption{\hspace{0.1cm}Qubit error rate (QBER) and frame error rate (FER) of an existing quantum LDPC code from \cite{Tan:2010} with length $N = 1034$ is illustrated.}
\vspace{-0.3cm}
\label{fig:SynVSSp}
\end{figure}
Fig.\ref{fig:SynVSSp} illustrates the performance  - \emph{i.e.} the qubit error rate (QBER) - our BP decoder achieves for an existing quantum stabilizer code, namely the code A (rate=1/2) of \cite{Tan:2010}. The lower solid line indicates the performance of our decoder with perfect channel information. Whereas, the upper solid line shows the performance of the decoder when no channel information has been made available at the decoder. The corresponding frame error rate (FER) is also shown for each case. Note, the QBER is the fraction of the qubits that possess an error, whereas the FER is the fraction of qubit blocks of length $N$ that contain at least one qubit error.


\subsection{Simulations for imperfect channel knowledge}
Consider now the case where a decoder can only attain partial channel information by probing the quantum channel using un-entangled and entangled quantum states (only one measurement each, \emph{i.e.} $N_m=1$). To simulate case $A$ and $B$ discussed in section \ref{QCE}, for each encoded block of qubits of length $N$ transmitted through the depolarizing channel, an estimated flip probability is randomly chosen at the receiver side from a normal distribution (truncated in a range of $0$ to $1$) characterized by mean $f^{'}$ and variance $(NJ(f^{'}))^{-1}$, where$f^{'} = 3f/4$.

\begin{figure}[htp]
\centering
\includegraphics[width=3.2in]{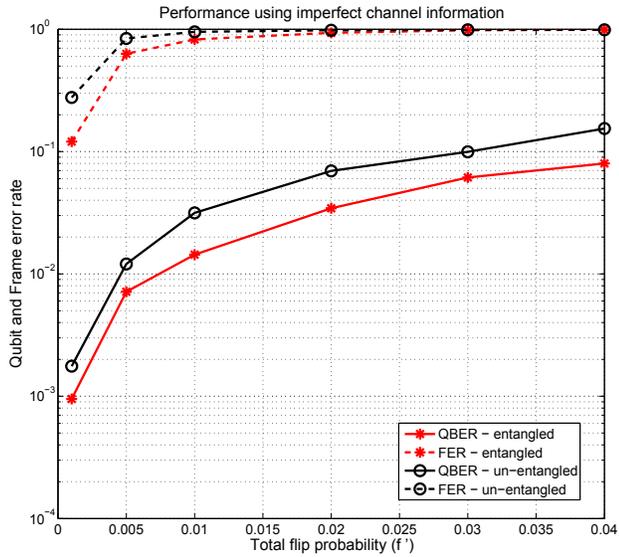}
\vspace{-0.3cm}
\caption{\hspace{0.1cm}Performance of a quantum stabilizer code of length $N = 1034$ with imperfect channel information estimated using entangled and un-entangled quantum states.}
\label{fig:ENvsUNE}
\end{figure}

Simulation results using imperfect channel knowledge as discussed above are illustrated in Fig.\ref{fig:ENvsUNE}. The quantum LDPC code used here is the same code adopted for Fig.\ref{fig:SynVSSp}. The decoding process terminates if $100$ block errors are collected or the maximum iteration number $(100)$ is reached. Contrary to Fig.\ref{fig:SynVSSp}, the results in Fig.\ref{fig:ENvsUNE} show the QBER and FER of the stabilizer code when only partial channel information on the channel is available at the decoder. We can clearly see that using entangled quantum states for estimation yields a better performance. Although not shown, as the mean value of $f^{'}$ approaches zero, the performance gap between the entangled and un-entangled estimation methods approaches the expected theoretical performance discussed in \ref{QCE}.

\begin{figure}[htp]
\centering
\includegraphics[width=3.2in]{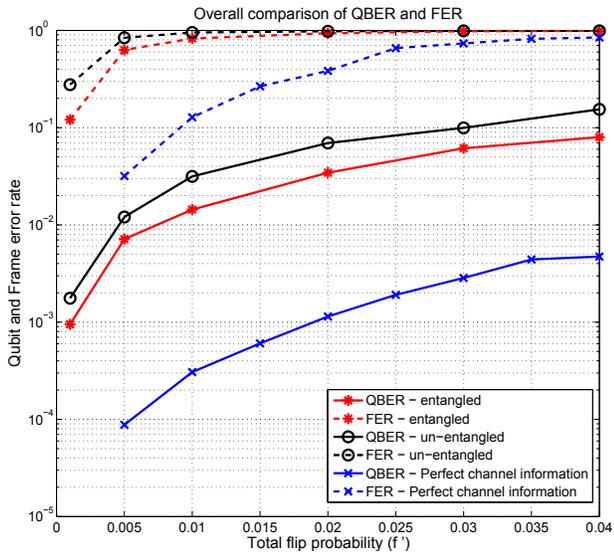}
\vspace{-0.3cm}
\caption{\hspace{0.1cm}Overall performance of the quantum stabilizer code with length $N = 1034$.}
\label{fig:Compare}
\end{figure}

In Fig.\ref{fig:Compare}, we collect the above results into the one plot for direct comparison purposes. Note that as the number of quantum states used to probe the channel goes to infinity, the performance of the code will approach to the performance of the curves marked as ``perfect channel information''. As we can see from the figure, even optimal channel identification using one probe measurement leads to roughly an order of magnitude hit on performance. We anticipate similar performance hits for any state-of-the-art quantum LDPC code. That the increase in the number of probing states leads to a better code performance, demonstrates the trade-off between the number of qubit probes (or entanglement consumption) and code performance.

It is important to note that, as is the normally the case, our performance curves are determined from simulation only. We have not developed any theoretical framework that can predict our new performance curves.  In this context we note the work of \cite{MHaga2010} in which LDPC code error performance is theoretically predicted for the binary symmetric channel as a function of a crossover probability $p_c$ and a different \emph{initialization} probability. It would be interesting to investigate whether extensions of the concepts introduced in \cite{MHaga2010} could deliver the necessary theoretical framework for directly predicting quantum LDPC code performance as a function of the quantum Fisher information on the channel parameters.

\section{Conclusion and future work}
 Utilizing quantum probes which are either in un-entangled  pure quantum states or maximally entangled qubit pairs, we have explored how channel identification in the quantum depolarizing channel affects the performance of quantum LDPC codes. Our results show the importance in quantum communications of a viable channel identification campaign prior to the decoding of any quantum codeword. Our work also highlights the trade-off between entanglement  consumption and quantum LDPC code performance. Future work will investigate similar issues for other quantum codes, consider more generic quantum channels, and pursue quantum LDPC code designs where uncertainty in the quantum channel is directly embedded.

\begin{thebibliography}{1}
\bibitem{Gallager:19633} R. G. Gallager, ``Low-density parity-check codes,'' \emph{PhD}, MIT, 1963.
\bibitem{MN1996} D. MacKay, ``Good error-correcting codes based on very sparse matrices,'' \emph{IEEE Transactions on Information Theory}, vol. 45, pp. 399-431, 1999.
\bibitem{Calderbank1996} A. R. Calderbank and P. W. Shor, ``Good quantum error-correcting codes exist,'' \emph{Phys. Rev. A}, vol. 54, pp. 1098-1105, 1996.
\bibitem{Steane} A. M. Steane, ``Error Correcting Codes in Quantum Theory,'' \emph{Phys. Rev. Letters}, vol. 77, pp. 793-797, 1996.
\bibitem{Gottesman1996} D. Gottesman, ``Class of quantum error-correcting codes saturating the quantum Hamming bound,'' \emph{Phys. Rev. A}, vol. 54, pp. 1862-1868, 1996.

\bibitem{Postol} M. S. Postol, ``A proposed quantum low density parity check code,''  \emph{arXiv:quant-ph/0108131v1}, 2001.
\bibitem{MacKay2004} D. MacKay, G. Mitchison, and P. McFadden, ``Sparse-graph codes for quantum error correction,'' \emph{IEEE Transactions on Information Theory}, vol. 50, pp. 2315-2330, 2004.
\bibitem{poulinturbo} D. Poulin, J. P. Tillich, and H. Ollivier, ``Quantum serial turbo codes,'' \emph{IEEE Transactions on Information Theory}, vol. 55,  pp. 2776-2798, 2009.
\bibitem{Nielsen:2000} M. Nielsen and I. Chuang, ``Quantum computation and quantum information,'' \emph{Cambridge Series on Information and the Natural Sciences}, Cambridge University Press, 2000.
\bibitem{Knill1997} E. Knill, R. Laflamme, A. Ashikhmin, H. Barnum, L. Viola and W. H. Zurek, ``Introduction to Quantum Error Correction,''  \emph{arXiv:quant-ph/0207170v1}, 2002.



\bibitem{Tan:2010} P. Tan and J. Li, ``Efficient quantum stabilizer codes: LDPC and LDPC-convolutional constructions,'' \emph{IEEE Transactions on Information Theory}, vol. 56, pp. 476-491, 2010.
    \bibitem{Hagiwara:2011} M. Hagiwara, K. Kasai, H. Imai, and K. Sakaniwa, ``Spatially coupled quasi-cyclic quantum LDPC codes,'' \emph{IEEE Proc. on International Symposium in Information Theory}, pp. 638-642, 2011.
        \bibitem{kenta2011} K. Kasai, M. Hagiwara, H. Imai, and K. Sakaniwa, ``Quantum Error Correction beyond the Bounded Distance Decoding Limit,'' \emph{IEEE Transactions on Information Theory}, \emph{arXiv:1007.1778v2 [cs.IT]}, 2011.

\bibitem{Kraus1983} Kraus, K., ``States, Effects and Operations: Fundamental Notions of Quantum Theory'', \emph{Lecture Notes in Physics}, vol. 190, Springer-Verlag, Heidelberg, Germany, 1983.
\bibitem{Fujiwara2001} A. Fujiwara, ``Quantum channel identification problem,'' \emph{Phys. Rev. A}, vol. 63,  042304, 2001.
\bibitem{Sasaki2002} M. Sasaki, M. Ban, and S. M. Barnett, ``Optimal parameter estimation of a depolarizing channel,'' \emph{Phys. Rev. A}, vol. 66,  022308, 2002.
\bibitem{Fujiwara2003} A. Fujiwara and H. Imai, ``Quantum parameter estimation of a generalized Pauli channel,'' \emph{Journal of Physics A: Mathematical and General}, vol. 36, no. 29, pp. 8093--8103, 2003.
\bibitem{Frey2010} M. R. Frey, A. L. Miller, L. K. Mentch, and J. Graham, ``Score operators of a qubit with applications,'' \emph{Quantum Information Processing}, vol. 9, pp. 629-641, 2010.
\bibitem{Frey2011} M. R. Frey, D. Collins, and K. Gerlach, ``Probing the qudit depolarizing channel,'' \emph{Journal of Physics A: Mathematical and Theoretical}, vol. 44, no. 20,  205306, 2011.
\bibitem{Poulin2008} D. Poulin and Y. Chung, ``On the iterative decoding of sparse quantum codes,'' \emph{Quantum Information Computation}, vol. 8,  pp. 987--1000, 2008.
\bibitem{MHaga2010} M. Hagiwara, M. P. C. Fossorier, H. Imai, ``LDPC Codes with Fixed Initialization Decoding over Binary Symmetric Channel'', \emph{IEEE Proc. on International Symposium in Information Theory}, pp. 784-788, 2010.

\end{thebibliography}
\end{document}